\documentclass[aps,prd,a4paper,nofootinbib,
preprintnumbers,superscriptaddress,showpacs,showkeys,twocolumn]{revtex4-1}

\usepackage{amsmath}
\usepackage{amssymb}
\usepackage{bm}
\usepackage{graphicx}
\usepackage{color}

\usepackage{amsfonts}
\usepackage{float}
\usepackage{placeins}
\usepackage{graphicx}
\usepackage[hidelinks]{hyperref}
\usepackage{color,amsmath,amssymb,bm}
\usepackage{tensor}

\usepackage[normalem]{ulem}

\newcommand{\be}{\begin{equation}}
\newcommand{\ee}{\end{equation}}
\newcommand{\ba}{\begin{eqnarray}}
\newcommand{\ea}{\end{eqnarray}}

\renewcommand{\sout}{\bgroup \color{red} \ULdepth=-.5ex \ULset}

\begin{document}

%\title{Heavy quark directed flow in small system in presence of electromagnetic fields}

 \title{Exploring the effects of electromagnetic fields and tilted bulk distribution on directed flow of D mesons in small systems}

\author{Yifeng Sun}
\affiliation{School of Physics and Astronomy, Shanghai Key Laboratory for Particle Physics and Cosmology,and Key Laboratory for Particle Astrophysics and Cosmology (MOE),
Shanghai Jiao Tong University, Shanghai 200240, China}

\author{Salvatore Plumari}
\affiliation{Department of Physics and Astronomy "E. Majorana", University of Catania, Via S. Sofia 64, 1-95123 Catania, Italy}
\affiliation{Laboratori Nazionali del Sud, INFN-LNS, Via S. Sofia 62, I-95123 Catania, Italy}

\author{Santosh K. Das}
\affiliation{School of Physical Sciences, Indian Institute of Technology Goa, Ponda-403401, Goa, India}

%\vspace{10pt}
%\begin{indented}
%\item[]February 2016
%\end{indented}

\begin{abstract}
We  studied the directed flow of  heavy quarks in small systems produced in p-Pb collisions due to both the impact of initial vorticity and electromagnetic fields. We employed a relativistic transport code to model the bulk evolution of the small systems and studied the heavy quark momentum evolution using Langevin dynamics. For the heavy quarks interaction with the bulk,  we employed a quasiparticle model (QPM). We observed a  large directed flow splitting ($\Delta v_1$) of charm quarks due to electromagnetic fields, which is comparable to  the directed flow splitting of charm quarks in nucleus-nucleus collisions. However, the magnitude of the directed flow due to the initial tilted matter distribution in p-nucleus collisions is not substantial. The observed directed flow is not rapidity odd due to the asymmetry in the colliding system. The results presented in this manuscript  provide an independent way to quantify the initial electromagnetic field produced and the matter distributed in small systems.
\end{abstract}



\maketitle

\section{Introduction}
A hot and  dense phase of nuclear matter, the quark-gluon plasma (QGP)~\cite{ Shuryak:2004cy, Science_Muller}, is
expected to  be produced  in the high-energy  nucleus-nucleus collisions at the Relativistic Heavy-Ion Collider (RHIC) and the Large Hadron Collider (LHC). Probing and characterizing  the bulk
properties of QGP is a field of high contemporary interest, and significant
progresses have been made in the last decades towards  understanding  the properties
of strongly interacting QGP. Currently, the description of the expanding dynamics in nucleus-nucleus collisions is quite robust, paving the way for new frontiers in ultrarelativistic heavy ion collisions (uRHICs) research. These frontiers include exploring the effects of  the largest relativistic vorticity and the strongest electromagnetic field ever created in a physical system.

Heavy quarks (HQs), mainly charm and bottom, are produced early due to their large masses. In fact, they are produced in the very early stage via hard scattering processes on a time scale $\tau=O(1/2m)$, where $m$ is the rest mass of the quark. Furthermore, due to their large relaxation time, heavy quarks are considered as  ideal probes~\cite{Prino:2016cni,Andronic:2015wma,Rapp:2018qla,Aarts:2016hap,Cao:2018ews,Dong:2019unq,Xu:2018gux,Svetitsky:1987gq, GolamMustafa:1997id,Uphoff:2011ad,Song:2015sfa,Cao:2016gvr,Plumari:2017ntm,
 Gossiaux:2008jv, rappv2, rappprl, Das:2010tj, Alberico:2011zy, Das:2013kea, Lang:2012cx, He:2012df, Xu:2017obm,Katz:2019fkc, Scardina:2017ipo, Song:2015ykw, Nahrgang:2014vza, Das:2016llg, Plumari:2019hzp,Sambataro:2022sns}  that  keep traces of
both the initial stage and the subsequent evolution into a thermalized QGP phase of heavy ion collisions.
Hence, heavy quarks are considered as novel probes to characterize the initial tilt of the QGP and electromagnetic fields~\cite{Das:2016cwd,Chatterjee:2017ahy,Oliva:2020doe, Oliva:2020mfr, Dubla:2020bdz, Sun:2020wkg, Chatterjee:2018lsx, Beraudo:2021ont,Jiang:2022uoe}, which can be probed perfectly through the directed flow ($v_1$) measurements. In a previous study~\cite{Das:2016cwd}, some of us predicted an electromagnetically-induced splitting in the directed flow of charm and anti-charm through  $D$ and  $\overline{D}$ meson, studied within the Langevin dynamics coupled with the Maxwell equations. The magnitude of the  directed flow splitting of $D$ mesons and their antiparticles is predicted to be an order of magnitude larger than that of the light charged hadrons. In another study, the authors investigated~\cite{Chatterjee:2017ahy}  $D$ meson directed flow, considering the tilt of the fireball in the reaction plane with respect to the beam axis. They predicted a larger rapidity odd directed flow of $D$ mesons in non-central heavy ion collisions compared to that of the light charged hadrons. 

Recently, both the STAR and ALICE  Collaborations have measured the  directed flow of $D$ mesons in experiments~\cite{STAR:2019clv,ALICE:2019sgg}.
Both  collaborations observed a non-zero directed flow for the $D^0$ meson. The absolute 
value of   the $D^0$ meson $dv_1/dy$ measured  by STAR Collaboration is about 25 times larger than that of the charged kaons.
The charge-dependent splitting in the  directed flow of heavy mesons at the 
highest RHIC energy is not clear and is smaller than the current precision
of the measurement. However, the ALICE Collaboration observed a positive slope for the splitting, 
which is about  3 orders of magnitude larger than  that of  the light charged hadrons.

 In recent years, there has been a significant effort to study collisions of small systems such as p-Pb and d-Au. One of the main questions is whether a strongly interacting QGP is formed in these systems. It has been observed that in high-multiplicity events of small systems, a strikingly collective behavior, a key observable used to gauge the production of the QGP,  is observed, similar to that in nucleus-nucleus collisions. However, measurements of the jet quenching and heavy quarkonium suppression show no evidence of the QGP. So far, only a few attempts have been made to study heavy quark dynamics in small systems produced in p-Pb collisions~\cite{Beraudo:2015wsd, Ruggieri:2018rzi, Liu:2019lac, Zhang:2020ayy,Haque:2021qka, Zhang:2022fum}. The influence of the electromagnetic fields and the tilt of the fireball in the small system on heavy quark dynamics and directed flow has not been explored yet. Therefore, a study on the directed flow of heavy quarks in small systems could be a good probe of the tilt of fireball and electromagnetic fields and could help answer the question of whether a strongly interacting QGP is formed in small systems.
%Although naively it is not expected that a QGP can be produced in these systems, however. It has also been shown that extremely intense electromagnetic fields can be produced in small systems as non-central heavy ion collisions, mainly due to the motion of spectator charges.} 

The paper is organized as follows: Sec. II discusses the initial distribution of bulk and charm quarks in p-Pb collisions. Sec. III presents the time evolution of electromagnetic fields in p-Pb collisions. Sec. IV presents the results on the average and splitting of the directed flow of $D$ and $\overline{D}$ mesons due to the tilt of the fireball and electromagnetic fields. Finally, a summary and conclusion are provided in Sec. V.  Throughout the text, all physical quantities related to rapidity are expressed in the nucleon-nucleon center-of-mass frame.

\section{Initial  distribution of bulk and charm quarks in proton-lead collisions}
We implement a commonly used  profile  that breaks longitudinally boost invariant to generate the initial density of p-Pb collisions:
\begin{equation}
\rho(\mathbf{x_{\perp}},\eta_s)=\rho_0\frac{W(\mathbf{x_{\perp}},\eta_s)}{W(0,0)}H(\eta_s),
\end{equation}
where $\mathbf{x_{\perp}}$ represents the transverse coordinate, $\eta_s=\frac{1}{2}\ln\frac{t+z}{t-z}$ is the spacetime rapidity, $\rho_0$ is the density at the centre of the fireball and $H(\eta_s)$ is a function that takes into account  a finite extension in rapidity and is given by
\begin{eqnarray}
H(\eta_s)=\exp{\bigg\{ -\frac{(|{\eta_s}|-\eta_{flat})^2}{2\sigma_\eta^2}\theta(|{\eta_s}|-\eta_{flat})\bigg\}}.
\end{eqnarray}
It corresponds to a central flat plateau of extension $2 \eta_{flat}$ beyond which the density decreases according to a Gaussian distribution with smearing $\sigma_{\eta}$. $W$ is the wounded weight function given by
\begin{equation}
W(\mathbf{x_{\perp}},\eta_s)=2(N_A(\mathbf{x_{\perp}})f_{-}(\eta_s)+N_B(\mathbf{x_{\perp}})f_{+}(\eta_s)),
\end{equation}
where $N_{A/B}$ are generated by the wounded quark model as described in Refs.~\cite{Bozek:2016kpf,Sun:2019gxg}, and $f_{+/-}(\eta_s)$ are given by
\begin{eqnarray}
f_{+/-}(\eta_s)=\left\{
\begin{array}{lll}
1/0,                                       &\, \,&  \eta_{s} < -\eta_{m}, \\ 
\frac{\pm \eta_{s}+\eta_{m}}{2\eta_{m}},   &\, \,& -\eta_{m} \leq \eta_{s} \leq\eta_{m},\\ 
0/1,                                       &\, \,&  \eta_{s} > \eta_{m}
\end{array}
\right.
\end{eqnarray}
We have chosen the values of $\eta_m=5.7$, $\eta_{flat}=2.5$ and $\sigma_{\eta}=2.5$ such that the resulting shape of $dN_{\rm{ch}}/d\eta$ matches the one measured by the ATLAS Collaboration~\cite{ATLAS:2015hkr}. This profile can account for more particles produced in the direction of nucleus, and for greater asymmetry in events with larger multiplicity.

%%\begin{figure}[t!]
%	\begin{center}
%		\includegraphics[scale=0.3]{deta.eps}
%	\end{center}
%	\caption{\label{Fig:deta}The pseudorapidity dependence of multiplicity $\frac{dN}%{d\eta}$ for different $\eta_m$ .}
%\end{figure}

%The pseudorapidity dependence of multiplicity $\frac{dN}{d\eta}$ for two different
%$\eta_m$ are shown in  Fig.~\ref{Fig:deta} in comparison with the available experimental %results. We can mimic the shape of the experimentally measured $\frac{dN}{d\eta}$ within 
%uncertainty. These profile accounts indicate the asymmetry in particle production in p-%nucleus collisions.  

Because heavy quarks are produced in hard scattering processes that follow binary collisions, the distribution of charm quarks is expected to be symmetric with respect to $\eta_s$. Additionally, the transverse coordinate distribution of charm quarks is generated according to the binary nucleon-nucleon collisions, while the initial transverse momentum distribution is obtained by the Fixed Order+Next-to-Leading Log (FONLL) QCD~\cite{Cacciari:2005rk,Cacciari:2012ny}. 

The evolution of the fireball is described by a relativistic transport Boltzmann equation solved at a fixed shear viscosity to entropy density ratio $\eta/s(T)$ \cite{Plumari:2012ep,Plumari:2015cfa,Scardina:2017ipo,Plumari:2019gwq,Sun:2019gxg,Plumari:2019hzp}. In this approach, the interaction between light quarks and gluons is tuned to a fixed value of $\eta/s(T)$ that is realized via locally computing the bulk cross section according to the Chapmann-Enskog approximation \cite{Plumari:2012ep}. In this way we simulate the evolution of the fluid in analogy to what is performed within hydrodynamics.

 \section{Time evolution of electromagnetic fields}
 
In this section, we determine the time and space profile of electromagnetic fields in the center-of-mass frame of p-Pb collisions. Without losing generality, we assume to the  configurations of p-Pb collisions that the proton moves along the negative $z$ directions and  is located at $x=y=0$ fm in the transverse plane, and Pb nucleus moves along the positive  $z$ directions and is located at $x=b$ and $y=0$ with b being the impact parameter. $t=0$ fm$/c$ is the time when proton and Pb nucleus fully overlap.

We adopt the methods used in Refs.~\cite{Gursoy:2014aka,Gursoy:2018yai} and compute the electromagnetic field at an arbitrary spacetime point ($t,z,\vec{x}_{\bot}$) produced by a point charge moving with a constant velocity $\vec{v}$ along the beam direction at location $\vec{x}_{\bot}^{\prime}$ in the transverse plane and  in a medium with a constant electric conductivity $\sigma$. We  express the spacetime point in terms of proper time $\tau=\sqrt{t^2-z^2}$, spacetime rapidity $\eta_s=\tanh^{-1}(z/t)$ and the rapidity of the point charge $y=\tanh^{-1}(v_z)$. The electromagnetic field can then be evaluated as:
\begin{eqnarray}
&&eB_y=\alpha \sinh(y)(x-x^{\prime})\frac{\frac{\sigma|\sinh(y)|}{2}\sqrt{\Delta}+1}{\Delta^{\frac{3}{2}}}e^A
\\&&eE_x=eB_y \coth(y),
\end{eqnarray}
where $\Delta \equiv \tau^2\sinh^2(y-\eta_s)+(\vec{x}_{\bot}-\vec{x}_{\bot}^{\prime})^2$ and $A\equiv \frac{\sigma}{2}(\tau\sinh{y}\sinh(y-\eta_s)-|\sinh(y)|\sqrt{\Delta})$.

%\begin{figure}[t!]
%	\begin{center}
%		\includegraphics[scale=0.25]{eBy.eps}
%	\end{center}
%	\caption{\label{Fig:elby} Variation of $B_y$ with time at different $\eta_s$ in 0-10\% p-Pb collisions at 5.02 TeV.}
%\end{figure}

To evaluate the electromagnetic field in p-Pb collisions, we need to sum together the electromagnetic fields generated by all protons in the proton and Pb nucleus. These include the spectator protons, meaning that at location $\vec{x}_{\bot}^{\prime}$ one finds either colliding proton or Pb nucleus but not both, and participant protons, meaning that  at location $\vec{x}_{\bot}^{\prime}$ one finds both colliding proton and Pb nucleus. For the spectator protons, their rapidity is the same as the beam rapidity, denoted by $Y_0$. For the participant protons, because they lose some energy in the collision, we use the empirical distribution~\cite{Kharzeev:1996sq} 
 \begin{eqnarray}
&&f(y)=\frac{a}{2\sinh(aY_0)}e^{ay}, (|y|<Y_0)
\end{eqnarray}    
where $a\approx1/2$ for + moving participants and $a\approx-1/2$ for - moving participants.  Regarding the charge distribution, it is assumed to be uniformly distributed within a sphere of radius 0.84 fm for colliding proton and 6.5 fm for colliding Pb nucleus, respectively.

%\begin{figure}[t!]
%	\begin{center}
%		\includegraphics[scale=0.25]{eEx.eps}
%	\end{center}
%	\caption{\label{Fig:elex} Variation of $E_x$ with time at different $\eta_s$ in 0-10\% p-Pb collisions at 5.02 TeV.}
%\end{figure}

\begin{figure}[h]
\centering
\includegraphics[width=0.95\linewidth]{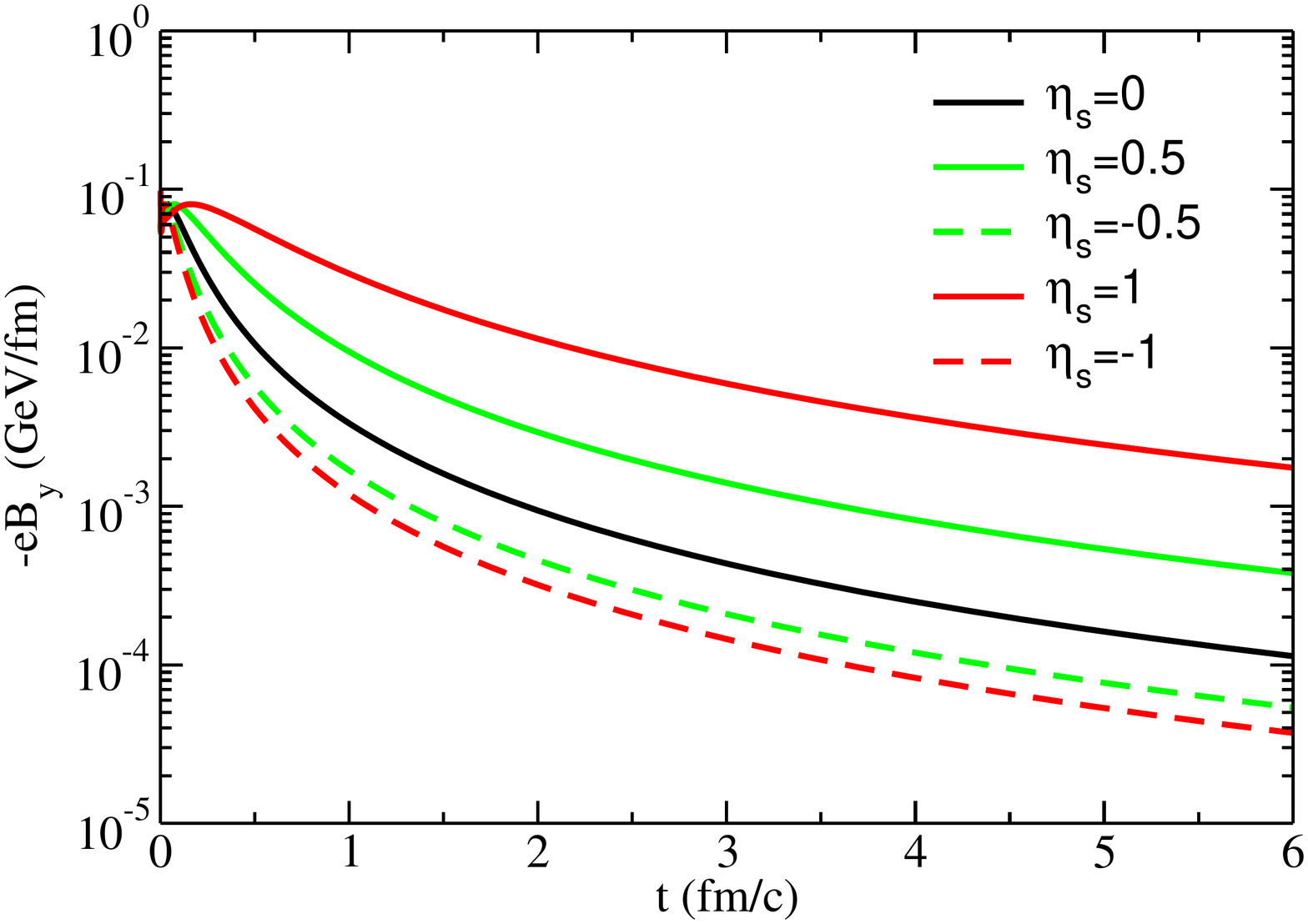}
\includegraphics[width=0.95\linewidth]{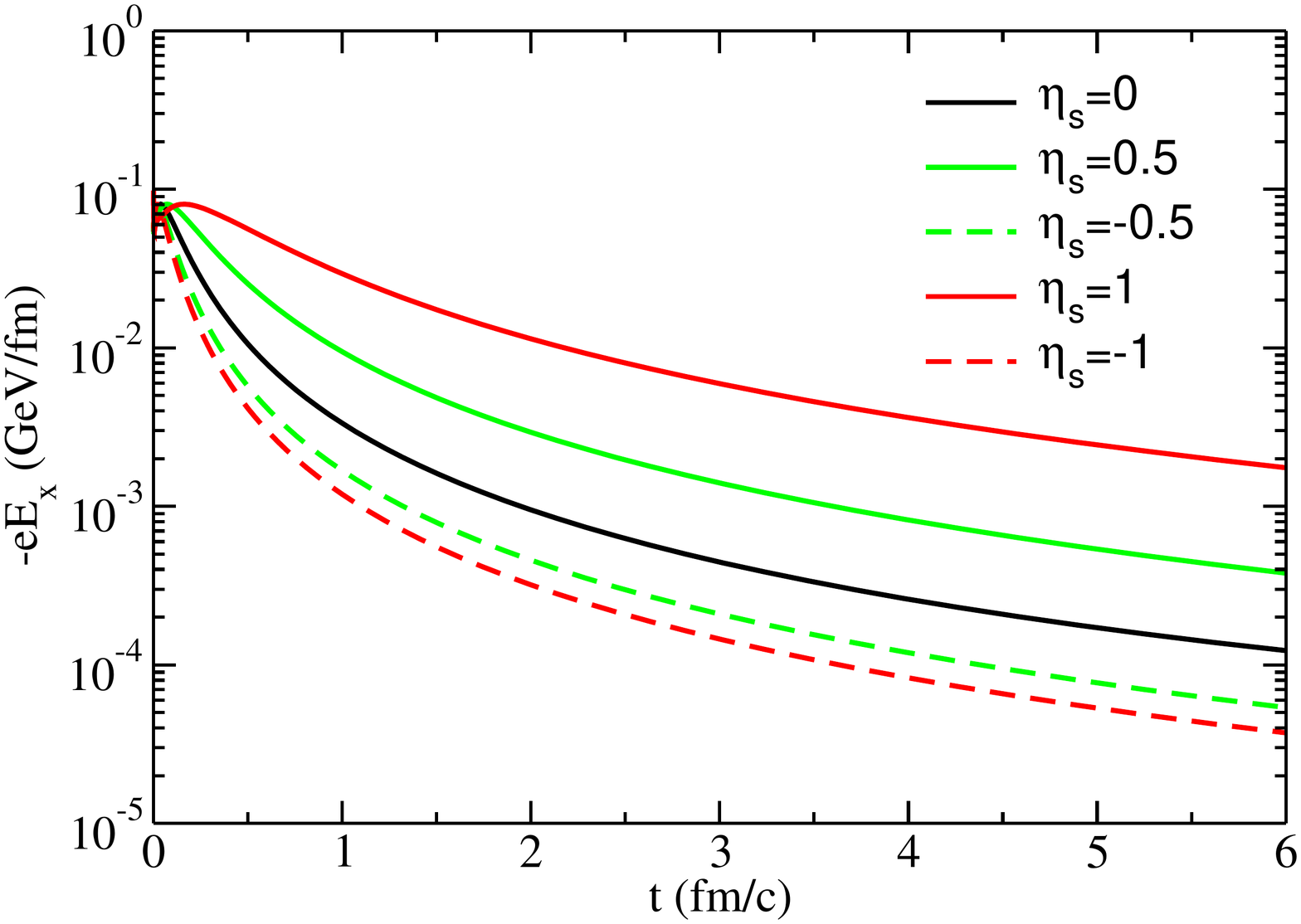}
\caption{Variation of $B_y$ (upper panel) and $E_x$ (lower panel) with time at different $\eta_s$ in 0-10\% p-Pb collisions at 5.02 TeV.}
\label{Fig:el}
\end{figure}

For 5.02 TeV p-Pb collisions at centrality 0-10\%, it is found $Y_0=8.58$ and the mean value of  the impact parameter is $\langle b\rangle=3.137$ fm. In Fig.~\ref{Fig:el}, we show the time evolution of electric and magnetic fields at $\vec{x}_{\bot}=0$ and different spacetime rapidity $\eta_s$ for 5.02 TeV p-Pb collisions at centrality 0-10\%, using the constant conductivity $\sigma_{el}=0.023$ fm$^{-1}$ within lQCD  calculations~\cite{Ding:2010ga,Amato:2013naa,Brandt:2012jc}. It is observed that the magnitudes of $E_x$ and $B_y$ are almost equal, and both are asymmetric with respect to $\eta_s$ and increase as $\eta_s$ increases.  This can be attributed to the fact that the electromagnetic fields in p-Pb collisions are primarily generated by the protons of Pb nuclei.

\section{Results}
The standard approach for studying the dynamics of heavy quarks in the QGP involves following their position and momentum evolution using the Langevin equation~\cite{rappv2,rappprl,Das:2010tj,Alberico:2011zy,Das:2013kea} or relativistic kinetic theory
\cite{Uphoff:2011ad,Gossiaux:2008jv, Song:2015sfa,Cao:2016gvr,Plumari:2017ntm,Das:2013kea,Scardina:2017ipo,Song:2015ykw,Nahrgang:2014vza, Berrehrah:2013mua, Das:2017dsh}. In this study, we employ the Langevin approach.
The momentum evolution of heavy quarks in QGP, with charge $q$ and momentum  $\mathbf{p}$, can be governed  by the Langevin equation in the presence of an external electromagnetic field, given by~\cite{Das:2016cwd}:
\begin{eqnarray}
 dx_i & = & \frac{p_i}{E}dt \ , \\
 dp_i & = & -\Gamma(p) p_i dt+C_{ij}(p)\rho_j\sqrt{dt} +{ F}_{i,ext}dt \ ,
 \label{lv1}\end{eqnarray}
where $dx_i$ and $dp_i$ are the changes of the coordinate and momentum in each discrete time step $dt$.
$\Gamma(p)$ and $C_{ij}(p)=\sqrt{2B_0}P_{ij}^{\perp}+\sqrt{2B_1}P_{ij}^{\parallel} $ are the drag force and covariance matrix. $P_{ij}^{\perp}= \delta_{ij}-p_i p_j/p^2$ and $P_{ij}^{\parallel}=p_i p_j/p^2$ are the transverse and longitudinal projector operators.
$B_0$ and $B_1$ are the transverse and longitudinal diffusion coefficients of heavy quarks.
At $p\rightarrow 0$, $B_0=B_1=D$,  $C_{ij}=\sqrt{2D(p)} \delta_{ij}$.  $\rho$ is the stochastic force with a
vanishing expectation value since there is no preferred
direction for the collisions.  $\Gamma(p)$ and $C_{ij}(p)$ that contain the physics of heavy quark drag and diffusion coefficients will encode all the information about the QGP medium. We have employed the fluctuation-dissipation theorem (FDT), $D=\Gamma E T$, where $T$ is the temperature of the thermal bath and $E$ is the energy of the heavy quark.
${F}_{i,ext}$ represents the external Lorentz
force due to the electromagnetic fields.

To compute the 
heavy quark transport coefficients, we employ a quasiparticle model (QPM)~\cite{  Das:2015ana, Plumari:2011mk}. In this study, we use the drag and diffusion coefficients that have been shown to reproduce the experimental measurements of
D meson observables~\cite{Scardina:2017ipo,Sun:2019fud,Plumari:2019hzp,Sambataro:2022sns} both at RHIC and LHC energies.  At the end of QGP phase, when the temperature of the bulk falls below the quark-hadron transition temperature of $T_c=155$ MeV, charm quarks are converted into $D$ mesons using the Peterson fragmentation function as done in Refs.~\cite{Scardina:2017ipo}.

\begin{figure}[t!]
	\begin{center}
		\includegraphics[scale=0.3]{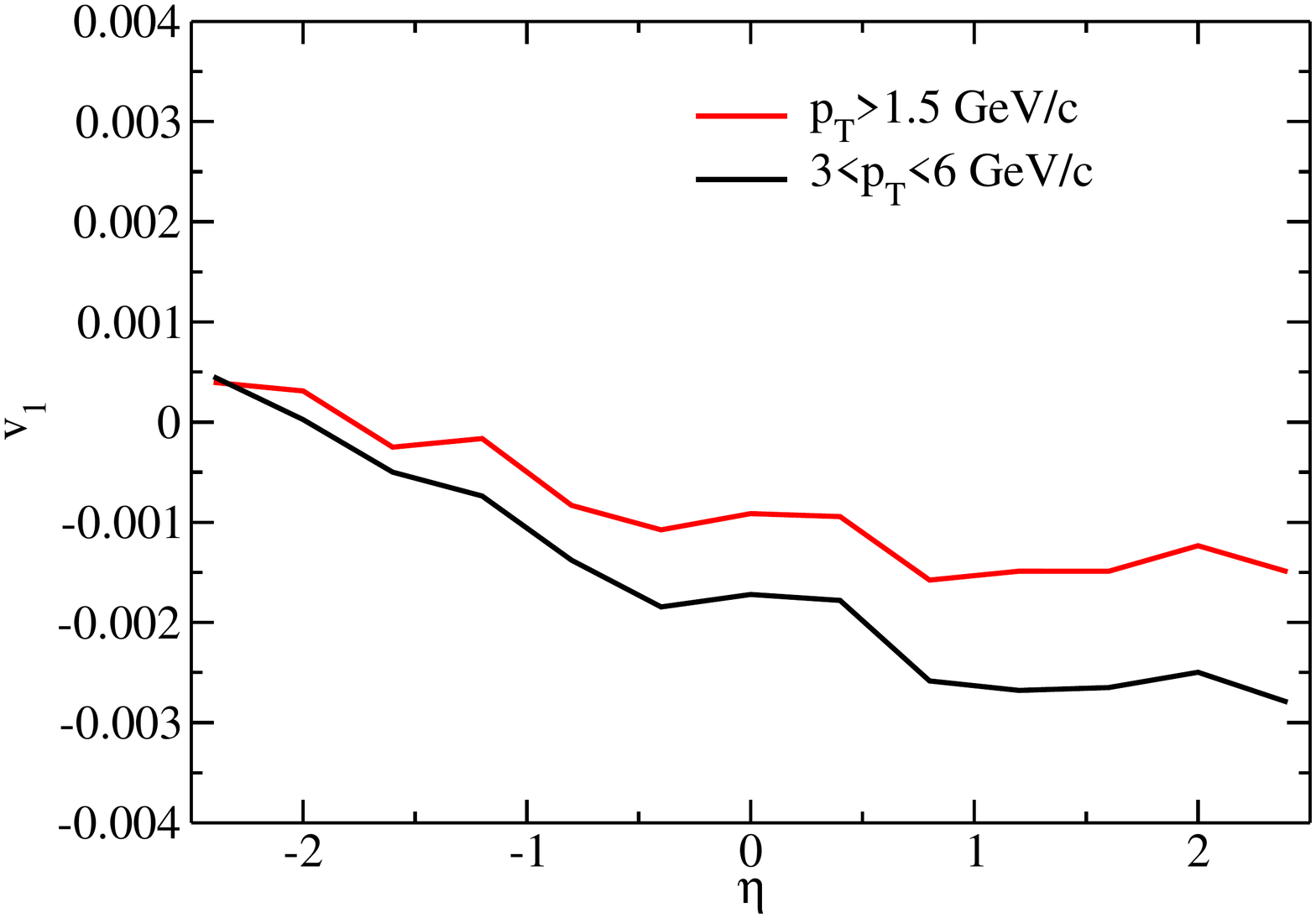}
	\end{center}
	\caption{\label{Fig:tilt} Directed flow $v_1$ of $D^0$ and $\overline{D}^0$ mesons  as a function $\eta$ at different $p_T$ cuts considering only the tilted distribution of the fireball in 0-10\% p-Pb collisions at 5.02 TeV..}
\end{figure}

The $v_1$ is defined as
\begin{eqnarray}
 v_1 & = & \langle\cos(\phi-\Psi_{RP})\rangle,
\end{eqnarray}
where $\phi$ denotes the azimuthal angle and $\Psi_{RP}$ is the reaction plane angle. Since the centers of both the proton and Pb nucleus  are alighed along the $x$ axis, the reaction plane angle is fixed at $\Psi_{RP}=0$ in our calculations.

In Fig.~\ref{Fig:tilt}, we present the variation of directed flow as a function of pseudorapidity $\eta$ obtained within Langevin dynamics at different $p_T$ cuts.  In this calculation, we have not included the electromagnetic field, so we obtained results considering only the 
initial tilted matter distribution. We observed a non-zero heavy quark directed flow 
due to the tilted initial distribution in p-nucleus collisions. The magnitude of this directed flow is larger for the $p_T$ range of $3 < p_T < 6$ GeV/$c$ compared to $p_T>$ 1.5 GeV/$c$, making it more applicable to be measured in experiments. Unlike in nucleus-nucleus collisions, the produced directed flow in p-nucleus collisions is not symmetric in pseudorapidity. However, the magnitude of the  directed flow in p-nucleus collisions is not substantial. This indicates the tilt in the produced matter distribution is not significant enough in  p-nucleus collisions to develop a sizable directed flow. Another reason for the small directed flow is  the shorter lifetime of QGP in p-nucleus collisions compared to nucleus-nucleus collisions.

\begin{figure}[t!]
	\begin{center}
		\includegraphics[scale=0.3]{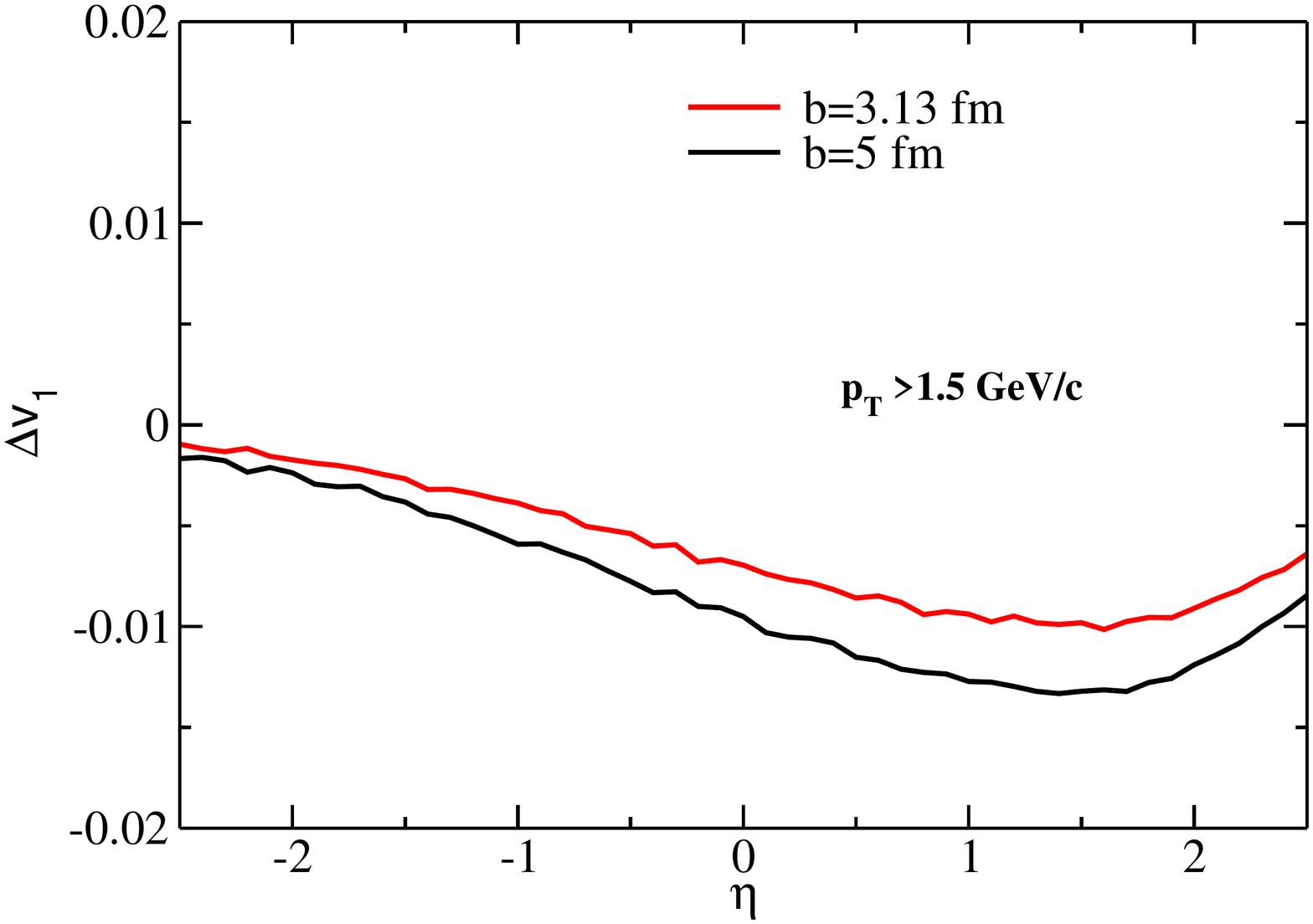}
        \includegraphics[scale=0.3]{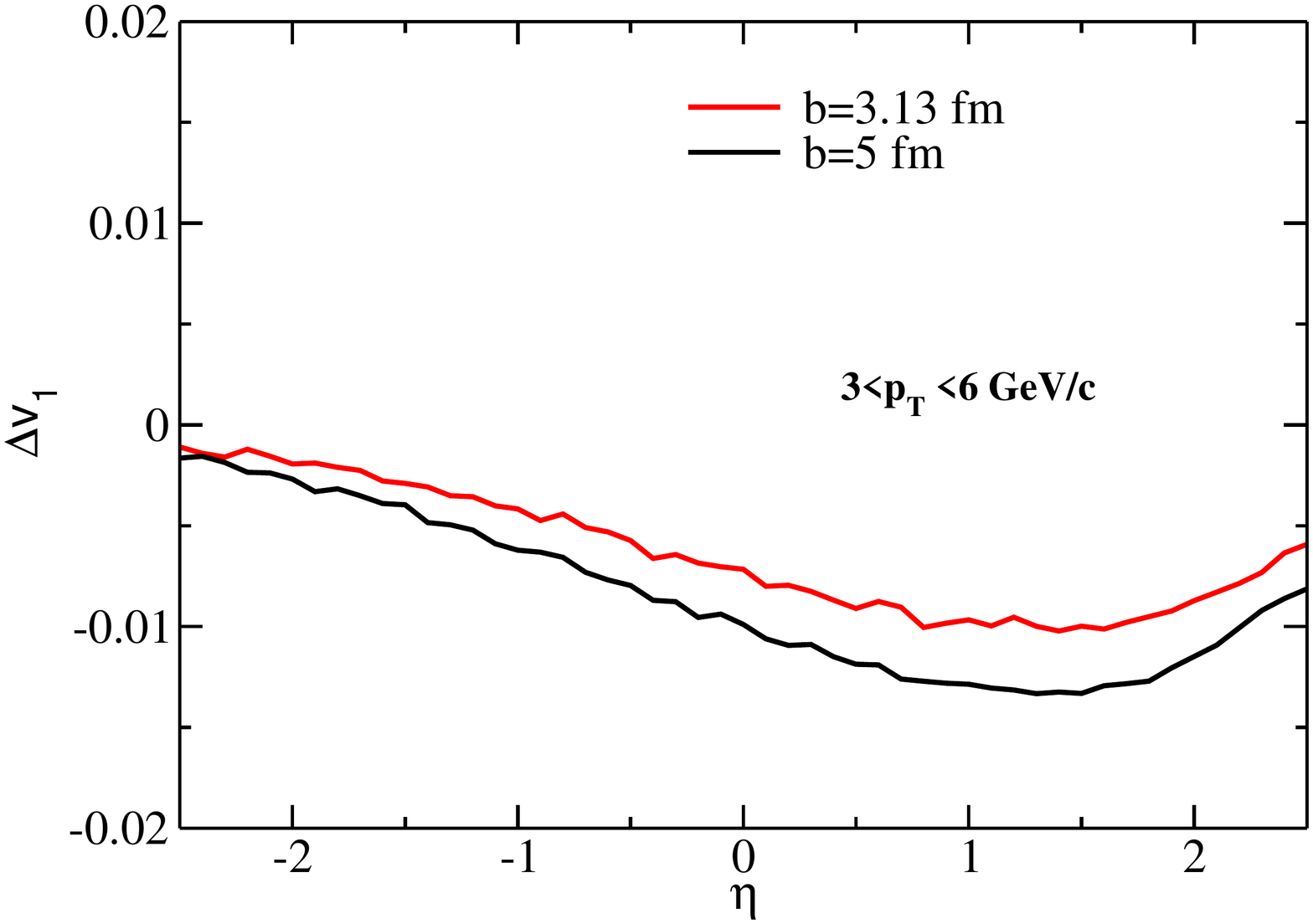}
	\end{center}
	\caption{\label{Fig:split1}Splitting in the directed flow, $\Delta v_1$, of $D^0$ and $\overline{D}^0$ mesons  at different $p_T$ cuts and impact parameters as a function of  pseudorapidity at 5.02 TeV p-Pb collision.}
\end{figure}

The splitting in the directed flow is considered as a signature of the strength of the produced electromagnetic fields. In Fig.~\ref{Fig:split1}, we show the variation of $\Delta v_1$ as a function of pseudorapidity in p-Pb collisions. The splitting is observed to be asymmetric in pseudorapidity, similar to the average directed flow, in p-Pb collisions. However, upon calculating different $p_T$ ranges of $3 < p_T < 6$ GeV/$c$ and $p_T>$ 1.5 GeV/$c$, the splitting of the directed flow is found to have a mild $p_T$ dependence compared to the average directed flow. Notably, the magnitude of the splitting due to the electromagnetic  field is substantial in a small system and is larger by about a factor 2 than one can expect from theoretical calculation in Pb-Pb collisions ~\cite{Oliva:2020doe}.

We notice that the splitting is sensitive to the impact parameter of the collisions. In Fig.~\ref{Fig:split1}, we  show as well the variation of $\Delta v_1$ as a function of pseudorapidity at different impact parameters b=3.13 and 5 fm. The impact of different impact parameters on the splitting is quite significant due to the increased strength of the electromagnetic fields at larger impact parameters.

\begin{figure}[t!]
	\begin{center}
		\includegraphics[scale=0.3]{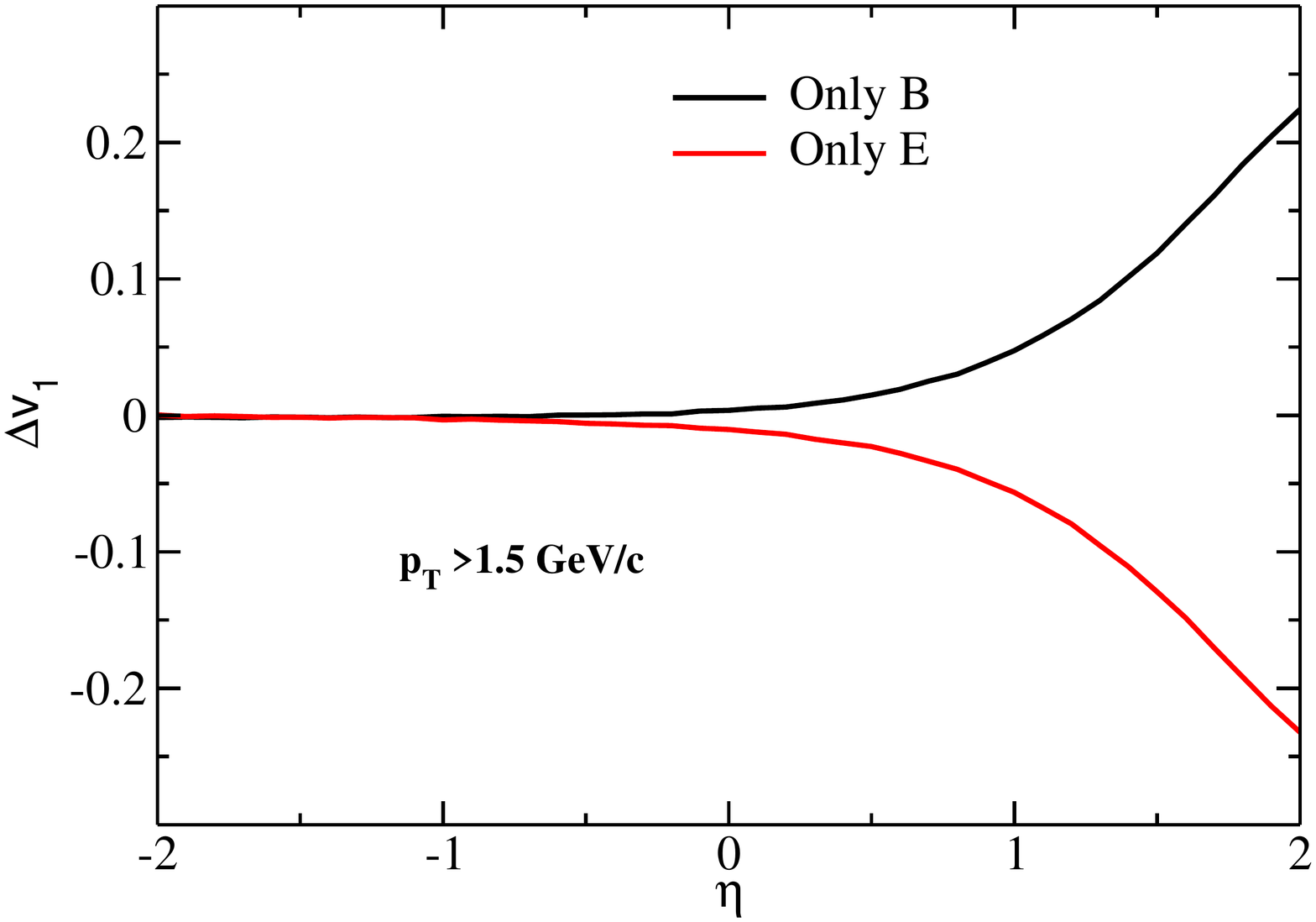}
	\end{center}
	\caption{\label{Fig:v1EB}Variation of the directed flow splitting, $\Delta v_1$, as a function of pseudorapidity considering only the electric or magnetic field in 0-10\% p-Pb collisions at 5.02 TeV.     }
\end{figure}

In Fig. 4, we also examine the variation of heavy quark directed flow splitting as a function of pseudorapidity considering only the electric or magnetic field, with the electric or magnetic field obtained using the same electric conductivity value of 0.023 fm$^{-1}$ taken from lQCD calculations. It is observed that the splittings increase strongly with $\eta$, due to the increase in the electric and magnetic fields' magnitude with $\eta_s$. At $\eta$=0, the directed flow splitting due to the electric field is non-zero, which is explained by the non-zero electric field at $\eta_s=0$. Additionally, the directed flow splitting due to the magnetic field is non-zero at $\eta=0$ due to the asymmetric distribution of the magnetic field. Combining the effects of both fields, the results indicate that the electric field dominates over the magnetic field, resulting in a negative splitting throughout the studied pseudorapidity range.

\section{Conclusions} 

In conclusion, we have studied the dynamics of heavy quarks in small systems produced in p-nucleus collisions, taking into account the impact of both electromagnetic fields and the initial tilted bulk distribution. The evolution of the fireball produced in the  p-nucleus collisions is described by a relativistic transport Boltzmann equation ~\cite{Sun:2019gxg} solved at a fixed shear viscosity to entropy density ratio which can reasonably describe the resulting shape of $dN_{\rm{ch}}/d\eta$ measured by the ATLAS Collaboration. To study the heavy quarks momentum evolution, we employed the Langevin equation. The impact of electromagnetic field on heavy quark dynamics is taken care of through the Lorentz force in the Langevin equation.  We adopt the methods used in Refs.~\cite{Gursoy:2014aka,Gursoy:2018yai} and compute the electromagnetic field at an arbitrary spacetime point  produced by a point charge moving with a constant velocity  along the beam direction at a given location  in the transverse plane and  in a medium with a constant electric conductivity $\sigma$.
The heavy quark bulk interaction is modeled within a quasiparticle model, which can describe the heavy quark  observables at nucleus-nucleus collisions both at RHIC and LHC energies.

We computed heavy quark directed flow developed in small systems produced in p-nucleus collisions. Our results show that the average directed flow of heavy quarks due to the tilted initial matter distribution in p-Pb collisions at 5.02 TeV is non-zero, although its magnitude is not substantial, unlike in Pb-Pb collisions. This indicates that the tilt in the produced matter distribution is not significant enough in p-nucleus collisions to
develop a sizable directed flow. However, the magnitude of the directed flow is quite sensitive to different $p_T$ cuts. Furthermore, we have observed a significant splitting in the directed flow of heavy quarks due to the electromagnetic fields in p-Pb collisions.
The magnitude of the observed splitting in the directed flow is larger than that of the splitting calculated in Pb-Pb collisions~\cite{Oliva:2020doe}.
It is also asymmetric in pseudorapidity and sensitive to the impact parameter of the collisions due to the increased strength of the electromagnetic fields at larger impact parameters. Recently similar conclusions have been reported for light hadron directed flow 
in the small system with electromagnetic fields~\cite{Oliva:2019kin}.   We notice that the electric field dominates over the magnetic field, resulting in a negative splitting throughout the studied pseudorapidity range. It is worth noting that the splitting is sensitive to the conductivity of the medium ~\cite{Oliva:2020doe}.

Our study highlights the importance of considering both the tilted bulk medium and the electromagnetic fields in understanding the heavy quark dynamics in small collision systems.  Heavy quark directed flow in a small system can provide an independent way to quantify the initial electromagnetic field produced and the matter distributed in small systems. For Pb-Pb collisions, the reaction plane or the spectator plane is reconstructed from spectator neutrons detected using the Zero Degree Calorimeter (ZDC) in ALICE. However, determining the reaction plane or the spectator plane for p-Pb collisions poses greater challenges due to the smaller resolution. Further efforts to the development of improved techniques or the exploration of alternative methods are needed in this regard.  The pre-equilibrium phase may also play an important role in a small system ~\cite{Sun:2019fud}. In this present study, we ignored the role of the pre-equilibrium phase. We will consider this  to compute the heavy quark observables in the small system in a forthcoming study.  \\

\begin{acknowledgements}
Y.S. thanks the sponsorship from Yangyang Development Fund.
S.K.D. acknowledges the support from DAE-BRNS, India, Project No. 57/14/02/2021-BRNS.
S.P. acknowledges the funding from UniCT under ‘Linea di intervento 3’ (HQsmall Grant).
\end{acknowledgements}

\end{document}